**Engineering the magnetic and magnetocaloric properties of PrVO$_3$ epitaxial oxide thin films by strain effects**

H. Bouhani[1,2,a)], A. Endichi[1,2], D. Kumar[3], O. Copie[1], H. Zaari[2], A. David[3], A. Fouchet[3], W. Prellier[3], O. Mounkachi[2], M. Balli[4], A. Benyoussef[2], A. El Kenz[2], and S. Mangin[1]

[1] Institut Jean Lamour, UMR 7198 CNRS-Université de Lorraine F-54000, France.

[2] Laboratory of Condensed Matter and Interdisciplinary Sciences (LaMCScI), Faculty of Science, Mohammed V University, 1014 Rabat, Morocco.

[3] Normandie Univ. ENSICAEN UNICAEN CNRS, CRISMAT 6 Boulevard Maréchal Juin, F-14050 Caen, Cedex 4, France.

[4] AMEEC Team, LERMA, ESIE, International University of Rabat, Parc Technopolis, Rocade de Rabat-Salé, 11100, Morocco.

**Abstract**

Combining multiple degrees of freedom in strongly-correlated materials such as transition-metal oxides would lead to fascinating magnetic and magnetocaloric features. Herein, the strain effects are used to markedly tailor the magnetic and magnetocaloric properties of PrVO$_3$ thin films. The selection of appropriate thickness and substrate enables us to dramatically decrease the coercive magnetic field from 2.4 T previously observed in sintered PVO$_3$ bulk to 0.05 T for compressive thin films making from the PrVO$_3$ compound a nearly soft magnet. This is associated with a marked enhancement of the magnetic moment and the magnetocaloric effect that reach unusual maximum values of roughly 4.86 $\mu_B$ and 56.8 J/kg K in the magnetic field change of 6 T applied in the sample plane at the cryogenic temperature range (3 K), respectively. This work strongly suggests that taking advantage of different degrees of freedom and the exploitation of multiple instabilities in a nanoscale regime is a promising strategy for unveiling unexpected phases accompanied by a large magnetocaloric effect in oxides.

[a)] **Authors to whom correspondence should be adressed**: hamza.bouhani@univ-lorraine.fr

The interplay between several degrees of freedom in complex functional materials gained a lot of interest due to its potential to enhance the caloric effects in new alternative cooling technologies such as magnetic cooling.[1-12] The latter which is based on the magnetocaloric effect (MCE) is an emergent, innovating and potentially low carbon technology. The MCE which is an intrinsic property of certain magnetic materials results in a change of their thermal state when subjected to an external magnetic field. Currently, the gadolinium metal (Gd) is the magnetic material used in the vast majority of magnetic cooling prototypes, mainly due to its magnetic phase transition taking place at 294 K leading to excellent magnetocaloric properties close to room temperature.[9, 13] However, this metal presents multiple disadvantages such as its easy oxidation as well as the limitation of its working temperature range close to the room temperature. In addition, gadolinium cannot be used in large scale applications because of its high cost. These issues have motivated the scientists to search for cheaper, safe and performant magnetocaloric materials under moderate magnetic fields including intermetallic and oxides.[1-11] The magnetocaloric effect in manganite-based perovskites exhibiting multiferroic behaviors have become an interesting topic because of the potential application of these oxides in some specific applications such as the liquefaction of hydrogen and space industry.[10-11] This sort of compounds fulfills the necessary conditions for practical applications as they unveil a strong chemical stability, high electrical resistivity, low hysteresis and mechanical stability.[1] In contrast, the magnetocaloric potential of the $RVO_3$ vanadates (R = rare earth) has not yet been explored except for the bulk $HoVO_3$.[14] However, perovskite-type vanadium oxides $RVO_3$ display a great variety of phase transitions associated with a series of charge, spin and orbital ordering phenomena making them interesting candidates from a magnetocaloric point of view.[14]

Today's research activities on magnetocaloric thin films attract a wide interest due to their potential integration in miniaturized electronic devices.[15, 16] This particularly motivated us to investigate the MCE in $RVO_3$ thin films since their behavior strongly depends on the cooperative nature of the Jahn-Teller distortion making them sensitive to strain effects including the induced biaxial strain due to the lattice mismatch between the substrate and the film. Such structural effects tend to play an important role in tuning the film properties.[17-18] In this work, we mainly focus on exploring and tuning the magnetic and magnetocaloric properties of high quality epitaxial $PrVO_3$ (PVO) thin films by applying a compressive strain via a proper choice of substrate.

The epitaxial PVO films were grown by pulsed-laser deposition (PLD) on two different cubic substrates, namely, (001)-oriented (La,Sr)(Al,Ta)$O_3$ (LSAT) and (001)-oriented $SrTiO_3$ (STO). Their in and out-of-plane lattice parameter are found to be 3.868 Å and 3.95 Å with a thickness of 41.7 nm for the LSAT film while those of the $SrTiO_3$ film are 3.905 Å and 3.92 Å with a thickness of 100 nm. The lattice parameters have been determined by perfoming X-ray diffraction measurements, see ref. [23] and [25] for details. The films deposition was carried out with a KrF excimer laser (λ = 248 nm) with a repetition rate of 2 Hz and a laser fluence of ∼ 2 J/$cm^2$ focusing on stoichiometric ceramic targets. The thin films were deposited at an optimum growth temperature ($T_{GROWTH}$) of 650 °C and under oxygen partial pressure ($P_{GROWTH}$) of $10^{-6}$ bar. In ZFC measurements our sample was cooled to the desired temperature under no magnetic field, then data were collected while heating under magnetic field. For the FC process, the sample was cooled in the presence of an external magnetic field to the desired temperature. We used two procedures, field cooled cooling (FCC) and field cooled warming (FCW) where data are collected during the cooling and heating processes, respectively. The magnetizations were performed by using the Quantum Design SQUID-VSM. Each hysteresis loop was measured after a 150 K excursion above the Néel temperature and corrected by subtracting the diamagnetic contribution arising from the substrate.

To investigate the strain effect on the magnetic properties of $PrVO_3$ thin films, the magnetization dependence on temperature was measured for PVO/$SrTiO_3$ under an applied magnetic field of 50 Oe as shown in Fig. 1. Hysteretic loop is also performed at 10 K in magnetic fields changing between –6 and 6 T (see the inset of Fig. 1). Magnetic measurements indicate hard-ferromagnetic behavior below 80 K being similar to that reported previously for bulk PVO.[19-21] In fact, the S-shape of magnetization depicts a metamagnetic transition which is defined as the transition between antiferromagnetic (AF) and ferromagnetic (F) configurations of spins under the effect of magnetic fields or temperature change.[22] The intrinsic coercivity is ∼ 2.8 T while the remanence magnetization is ∼ 48 emu/$cm^3$. The magnetization saturation is found to be ∼ 54 emu/cm³ being equivalent to only 0.291 $\mu_B$/ f.u at 10 K.

The presence of soft component can be seen at a magnetic field of ∼ 0.2 T indicated by the shape of M vs H loop caused by the magnetic field induced transition which is absent at higher temperatures as earlier observed.[23] The high coercivity may arise from the pinning mechanism due to the microstructure as well as the film unit cell orientation compared to the in-plane and out-of plane magnetic field directions. On the other hand, we observed a reduction of $T_N$ (see figure S2) compared to bulk PVO ($T_N \approx 140$ K) which could be explained by the oxygen vacancy-induced film lattice distortion.[24] XRD reveals that the pseudo cubic volume of PVO unit cell when deposited on $SrTiO_3$ (≃ 60.91 Å³) is larger than its equivalent of the bulk (≃ 58.86 Å³).[25] As a result, the volume expansion decreases the transfer integral which tends to reduce the neighbor exchange interactions as the magnetic interactions in this system are governed by super exchange mechanisms. It is worthy to mention that no significant magnetic anisotropy is observed when comparing the performed magnetic measurements under magnetic fields applied in and out of the sample plane. This can be related to the crystallographic orientation as a strong perpendicular magnetic anisotropy is revealed when the substrate orientation is changed from (001)- to a (111)- or a (110)-oriented $SrTiO_3$.[26]

Figures. 2(c)-(d) display some selected isothermal magnetization curves for two different orientations measured up to 6 T of a PVO film deposited on a (001)-oriented LSAT substrate. As shown, the magnetization saturation is markedly enhanced when compared to PVO/$SrTiO_3$, reaching about $\simeq$ 900 emu/$cm^3$ and $\simeq$ 785 emu/$cm^3$ at 3 K for $H\perp$ (Fig. 2(d)) and H// (Fig. 2(c)), respectively. At 10 K, the corresponding magnetization saturations are about $\simeq$ 402 emu/$cm^3$ and $\simeq$ 305 emu/$cm^3$, respectively. These values are much larger than those of $PVO/SrTiO_3$ as can be clearly seen from the inset of Fig. 1. In addition, the coercive field is largely reduced to attain about 0.3 T and 1.1 T for hysteretic loops performed in plane and out of plane (Fig. 2), respectively. More surprisingly, the coercive field decreases dramatically at 3 K reaching only 0.05 T for magnetic fields applied within the films plane as shown in Fig. 2(c). This markedly contrasts with the conventional magnets in which usually the thermal excitations lead to the reduction of coercivity. This contrast could be attributed to FM and AFM couplings or/and the spin and orbital transitions usually leading to stair-like hysteresis observed in bulk $PrVO_3$.[20] The enhancement of coercivity when heating may also be attributed to the stress induced magnetic anisotropy due to the relaxation of the surface stress reported for various magnetic thin films.[26] The temperature dependence of magnetization at an in-plane applied magnetic field of 50 Oe is displayed in Fig. 2(a). As can be clearly observed, a sharp decrease of magnetization at low temperature and a magnetic transition from paramagnetic (PM) to an antiferromagnetic (AFM) phase transition occurs at $T_N = 125$ K. Such a transition is attributed to the beginning of a G-type spin ordering (G-SO).[21] The differentiation of the temperature-dependent magnetization is displayed in the inset of Fig. 2(a) where two additional magnetic transitions take place at $T_2 \simeq 20$ K and $T_3 \simeq 80$ K. These transitions were absent in bulk $PVO$[27] but they were recently reported in strained $PVO$ films[20] and in doped $Pr_{1-x}Ca_xVO_3$ compounds.[28] Upon cooling down to 3 K, the plot of the first derivative of the magnetization temperature dependence exhibits a minimum at very low temperature which could be explained by the polarization of the praseodymium magnetic moments. The newly established order is due to the fact that the antiferromagnetic vanadium sublattice produces an exchange field that results in a ferrimagnetic structure of Pr sublattice under cooling as already observed in other vanadates.[28] This is supported by the presence of a soft component at temperatures below 20 K (Figs. 2(c)-(d)) and also by the fact that the magnetization saturation reaches 4.86 and 5.54 $\mu_B$ at 3 K when a magnetic field is applied in and out-of-plane, respectively. On the other hand, these findings inform us on the contribution of both $Pr^{3+}$ and $V^{3+}$ ions to the whole magnetization since the theoretical saturated moment of free $Pr^{3+}$ and $V^{3+}$ ions are 3.22 and 2.12 $\mu_B$ suggesting that all the praseodymium and vanadium moments are fully aligned parallel to the magnetic field.

The ZFC, FCW and FCC curves were measured in 50 Oe field applied in the sample plane from 3 to 300 K for $PVO/LSAT$ as shown in Fig. 2(b). The bifurcation between FC and ZFC magnetizations indicates an intrinsic disorder and irreversibility being the characteristic of a complex system. This difference reflects the impact of the anisotropy on the shapes of ZFC and FC curves below the ordering temperature since the coercivity is related to the magnetic anisotropy. The latter plays an important role in determining the magnetization at a given field strength during both the ZFC and FC processes since it aligns the spins in a preferred direction. During the ZFC process, the spins are locked in random directions since no magnetic field is applied while cooling the thin films to the desired temperature. When a small magnetic field is applied at temperatures far below $T_N$ and as the system is anisotropic,[29] the magnetization decreases to reach negative values indicating a possible competition between antiferromagnetic interactions, a characteristic which is observed in orthovanadate $RVO_3$ compounds.[30] A small negative trapped field in the sample space as well as the coercivity could be responsible for the negative magnetization.[31] During the FC process, the $PVO$ film is cooled under the application of a magnetic field. Therefore, the spins will be aligned in a specific direction depending on the strength of the applied magnetic field. Consequently, $M_{FC}$ continuously increases below $T_N$ as the temperature decreases.

Magnetic isotherms collected under magnetic fields going from 0 up to 6 T at different temperatures are reported in Figs. 3(a)-(b) for the $PVO/LSAT$ films. Except the isothermal magnetization at 3 K which shows a typical behavior of a ferromagnetic material, all the other isotherms follow a sharp increase when the magnetic field is below 30 kOe indicating a field induced first order metamagnetic transition from AFM to FM state as a result of the strong competition between Pr 4f and V 3d spins.[23] Such a competition often leads to a giant MCE in strongly correlated materials.[32-34] A similar behavior is found in the corresponding Arrot plots (M² versus H/M) confirming the first order nature of the transition according to Banerjee[35] criterion as the curves show negative slope at some points (not shown here).

The large field-induced metamagnetic transition in $PVO/LSAT$ films and the soft component in M vs H below 30 K are a clear indication of a possible giant magnetic entropy change. In order to explore the magnetocaloric effect in $PVO$ films, magnetic field-induced entropy change $-\Delta S_M$ was calcuted from magnetic isotherms by using the well-known Maxwell relation (MR) given as follow[1]:

$$\Delta S_M = \int_0^H \left(\frac{\partial M}{\partial T}\right)_H dH$$

We are aware that the utilization of Maxell relation to evaluate entropy changes in materials showing large hysteretic effects could lead to spurious values as already demonstrated by one of the present paper authors.[36] However, the Maxwell relation can be also used to reasonably evaluate the MCE in term of the entropy change even in first-order phase transition materials provided that the remaining magnetization from previous isotherms is suppressed via a thermal loop.[37] On the other hand, it has been demonstrated recently that the magnetocaloric effect in multiferroics could be evaluated perfectly via Maxwell relation.[38] In this way, it has been particularly found that the deduced entropy change from magnetization measurements of $EuTiO_3$ fits perfectly with that obtained from specific heat data.[38] For more information about the impact of hysteretic phenomena on the MCE, we refer the reader to Refs. 34 and 35.

In our case, magnetic isotherms of Figs. 3(a)-(b) are used to calculate the entropy change exhibited by the $PVO$ films on $LSAT$. However, since the entropy change is proportional to the area between two successive isotherms, $\Delta S_M$ was directly calculated without subtracting the magnetic contribution arising from the substrate as already done in the case of $La_2NiMnO_6$ thin films.[39] For $PVO/STO$, it was difficult to calculate the MCE because of the overlap between M vs H curves as well as the very low (negligible) magnetization (see supplementary materials) at low temperatures.

The temperature dependence of the magnetic entropy change unveils larger values at very low temperature for $PVO/LSAT$ films. $-\Delta S_M$ reaches roughly a maximum value of 56.7 $J\ kg^{-1}\ K^{-1}$ for a magnetic field changing from 0 to 6 T applied in the sample plane (Fig. 3(a)), being about 63% of its theoretical limit given by R*Ln(2J+1). In a similar field change applied out of plane (Fig. 3(d)), $-\Delta S_M$ is slightly lower and found to be about 52.7 $J\ kg^{-1}\ K^{-1}$. Also, the magnetic entropy change shows a large magnetocaloric effect under relatively low magnetic fields that can be easily reached via permanent magnets. In the magnetic field change of 2 T applied within and out of thin films plane, $-\Delta S_M$ reaches 19.5 and 16.3 $J\ kg^{-1}\ K^{-1}$, respectively. As shown in Figs. 3(c)-(d), a large magnetocaloric effect can be induced below the AFM transition temperature. This suggests that a major part of the contribution to the MCE comes from the praseodymium 4f spins. A list of some relevant magnetocaloric materials working in the cryogenic temperature range are reported in Table 1 for comparison. As shown, the exhibited $\Delta S_M$ by the strained $PVO$ film is significantly larger than its equivalent reported for several rare-earth metal transition oxides making the PVO films potential candidates for low temperature magnetic refrigeration.

To sum up, we have investigated the magnetic and magnetocaloric properties of $PVO$ films grown by pulsed laser deposition, in view of their potential application in cryogenic magnetic cooling. The obtained results reveal that the magnetic and magnetocaloric properties of $PVO$ compounds can be easily tailored by using the thin films approach. Particularly, the coercive magnetic field was dramatically decreased making from the $PVO$ compound a nearly soft magnet. Accordingly, a giant MCE is exhibited by $PVO$ thin films grown on $LSAT$ substrates at low temperatures pointing out the great impact of strain effects and the competition between AFM and FM exchange interactions. These finding would open the way for the implementation of $PVO$ thin films in some specific applications such as on-chip magnetic micro-refrigeration and sensor technology. Our result not only suggests that epitaxial $PVO$ thin films present a non-negligible potential for refrigeration at cryogenic temperatures but may also pave the way for new applications taking for example advantage of the possibility to tailor their magnetic coercivity. However, we are aware that the reported entropy change values in $PVO$ thin films are too large when compared to the best magnetocaloric materials working in a similar working temperature range. The necessary was done to reasonably evaluate the MCE in term of the entropy change by considering the impact of hysteretic phenomena. In order to accurately estimate $\Delta S$, the measurements of specific heat in equilibrium conditions are highly required. This point will be certainly addressed in the future.

See the supplementary material for in-plane MH loops for PVO/STO film as well as its differentiation of the temperature-dependent magnetization, and higher temperatures MH loops for PVO/LSAT film with magnetic field applied in and out-of-plane.

**DATA AVAILABILITY**

The data that support the findings of this study are available from the corresponding author upon reasonable request.

**ACKNOWLEDGMENTS**

This work was supported by PHC Toubkal 17/49 project. M. Balli highly appreciates the financial support from the International University of Rabat.

**Fig captions**

**Fig. 1** Magnetization dependence of temperature for $PrVO_3$ film on $SrTiO_3$ substrate performed under an in-plane applied magnetic field of 50 Oe. Inset displays the magnetic hysteresis loops measured at 10 K after subtracting the diamagnetic contribution of the substrate and the holder.

**Fig. 2** (a) Temperature dependence of magnetization of $PrVO_3$ on LSAT thin film under an in-plane applied magnetic field of 50 Oe. Inset: differentiation of the temperature-dependent magnetization. (b) Temperature dependence of magnetization in zero field-cooling (ZFC), field cooled cooling (FCC) and field cooled warming (FCW) conditions. (c, d) Some selected magnetic isotherms in the temperature range 3-13 K for PVO/LSAT collected in-plane (c) and out-of-plane (d) applied magnetic field.

**Fig. 3** Magnetic and MCE properties of $PrVO_3$ on LSAT. a) Magnetic isotherms in the temperature range of 3-31 K with a step of 2 K under an in-plane (a) and out-of-plane (b) magnetic fields. (c, d) Temperature dependence of magnetic entropy change of $PrVO_3$/LSAT under some selected magnetic fields applied within (c) and out-of-plane (d).

**Tables**

**Table 1.** Maximum magnetic entropy change $\Delta S_M$ shown by PrV$O_3$ deposited on LSAT substrate compared to some relevant cryomagnetocaloric compounds. SC means single crystal

| Materials | T(K) | ΔH (T) | ΔS(J/kg K) | Ref. |
|---|---|---|---|---|
| PVO/LSAT (//-⊥) | 3 | 6 T | 56.7-52.7 | Present work |
| GdFeO$_3$ (SC) | 2.5 | 6 T | 43.1 | 40 |
| EuTi$O_3$ | 5.6 | 5 T | 42.4 | 38 |
| HoV$O_3$ (SC) | 15 | 7 T | 17.2 | 14 |

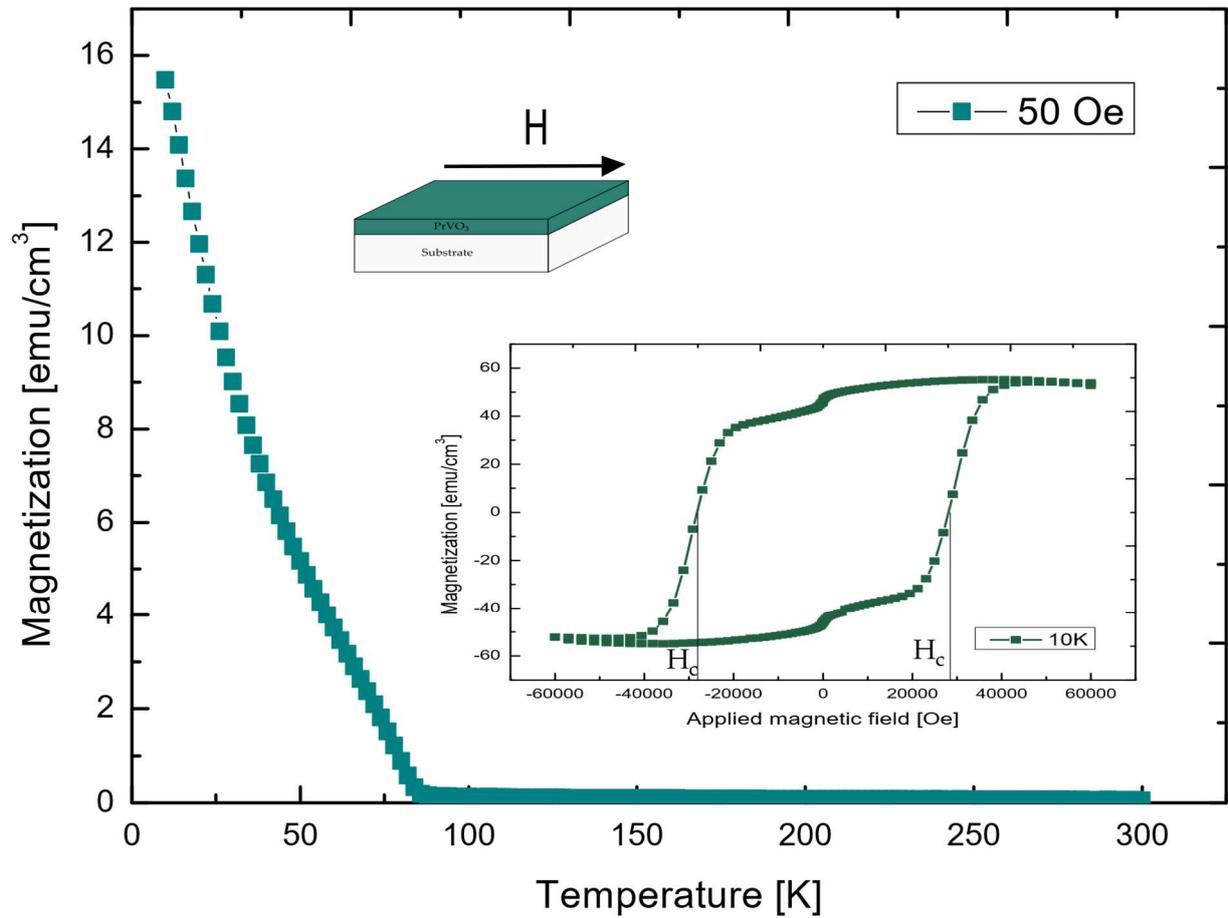

**Figure 1.** Magnetization dependence of temperature for PrV$O_3$ film on SrTi$O_3$ substrate performed under an in-plane applied magnetic field of 50 Oe. Inset displays the magnetic hysteresis loops measured at 10 K after subtracting the diamagnetic contribution of the substrate and the holder.

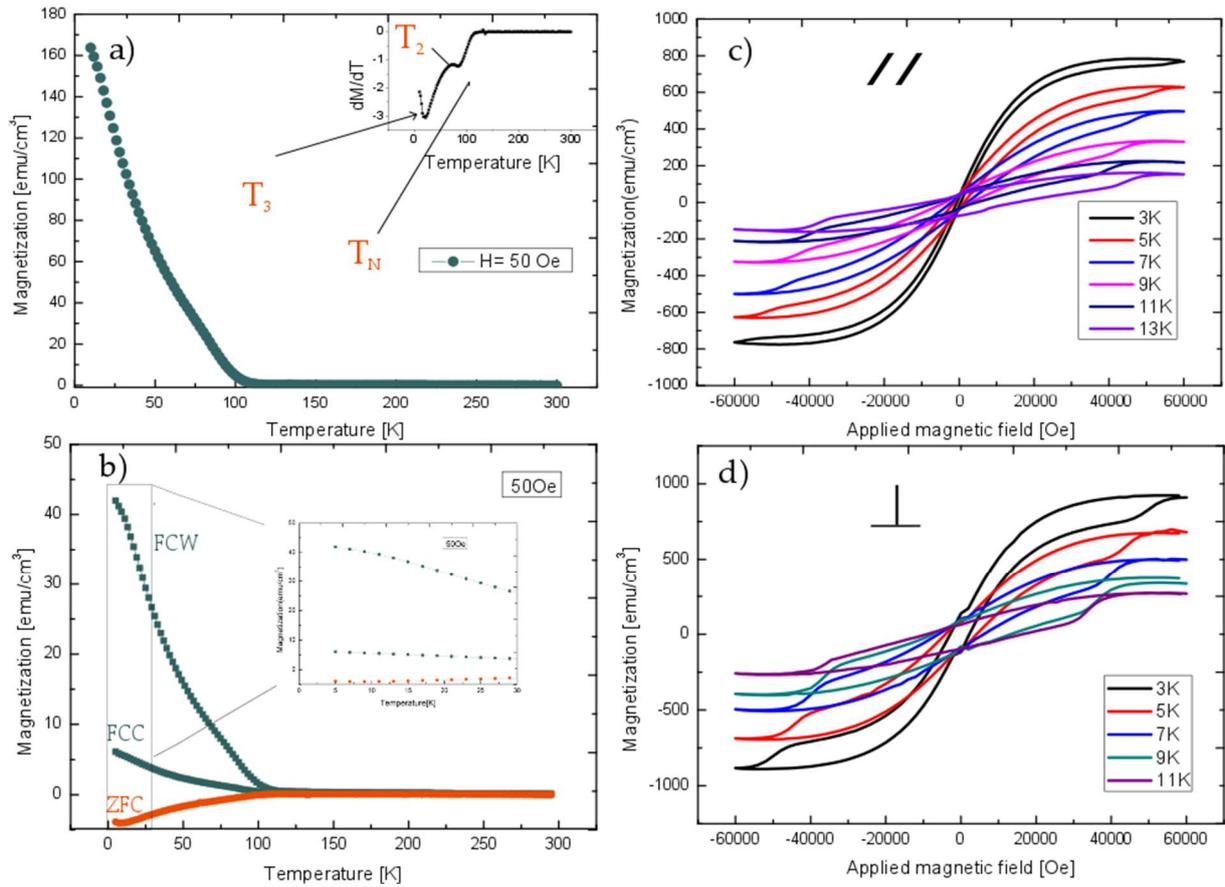

**Figure 2.** (a) Temperature dependence of magnetization of PrV$O_3$ on LSAT thin film under an in-plane applied magnetic field of 50 Oe. Inset: differentiation of the temperature-dependent magnetization. (b) Temperature dependence of magnetization in zero field-cooling (ZFC), field cooled cooling (FCC) and field cooled warming (FCW) conditions. (c, d) Some selected magnetic isotherms in the temperature range 3-13 K for PVO/LSAT collected in-plane (c) and out-of-plane (d) applied magnetic field.

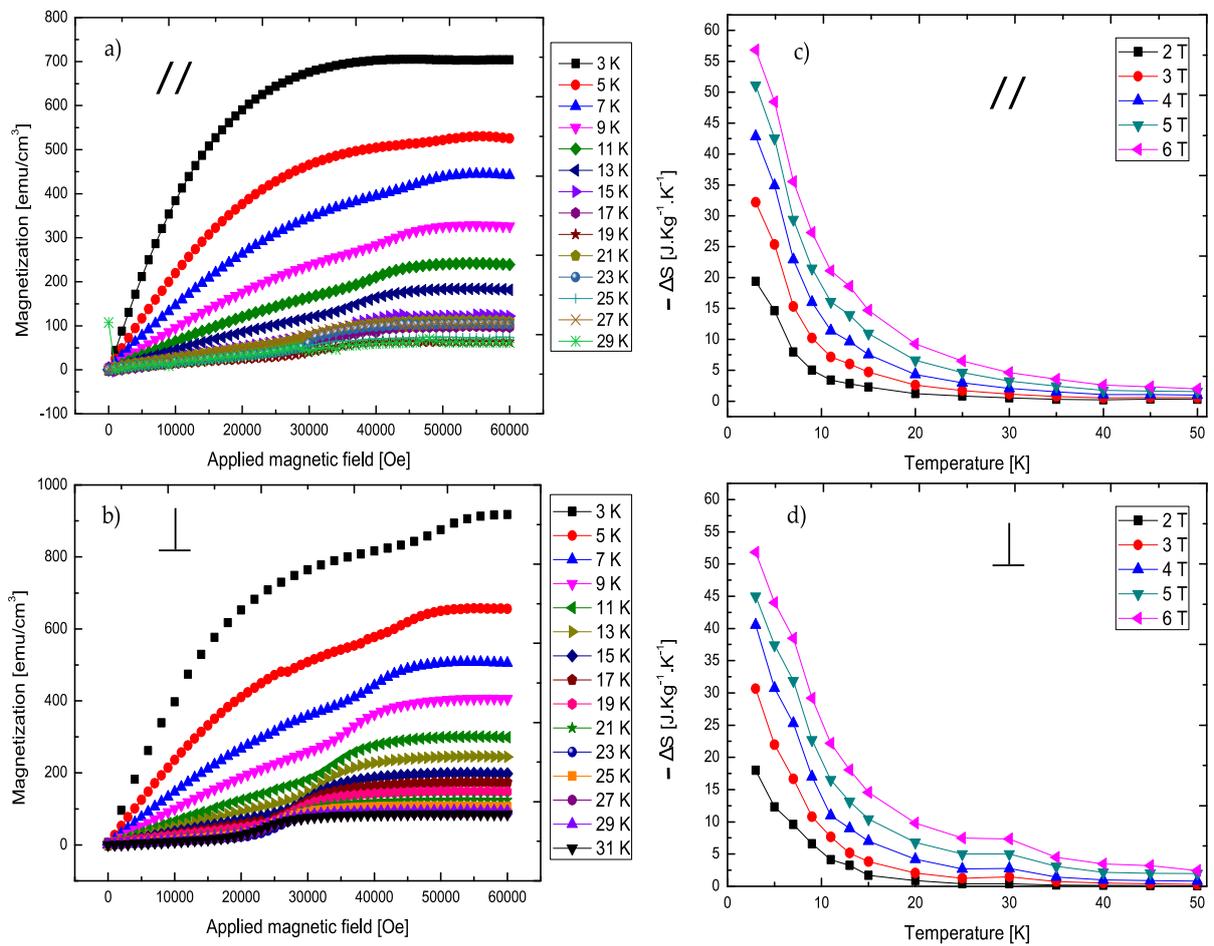

**Figure 3.** Magnetic and MCE properties of PrV$O_3$ on LSAT. a) Magnetic isotherms in the temperature range of 3-31 K with a step of 2 K under an in-plane (a) and out-of-plane (b) magnetic fields. (c, d) Temperature dependence of magnetic entropy change of PrV$O_3$/LSAT under some selected magnetic fields applied within (c) and out-of-plane (d)

14